\documentclass[12pt]{article}

\usepackage{hyperref}
\usepackage{graphicx}

\title{Spectral statistics of random geometric graphs}

\author{C. P. Dettmann \and O. Georgiou \and G. Knight}

\begin{document}

\maketitle

\begin{abstract}
We use random matrix theory to study the spectrum of random geometric graphs, a fundamental model of spatial networks. Considering ensembles of random geometric graphs we look at short range correlations in the level spacings of the spectrum via the nearest neighbour and next nearest neighbour spacing distribution and long range correlations via the spectral rigidity $\Delta_3$ statistic. These correlations in the level spacings give information about localisation of eigenvectors, level of community structure and the level of randomness within the networks. We find a parameter dependent transition between Poisson and Gaussian orthogonal ensemble statistics. That is the spectral statistics of spatial random geometric graphs fits the universality of random matrix theory found in other models such as Erd\H{o}s-R\'{e}nyi, Barab{\'a}si-Albert and Watts-Strogatz random graphs.
\end{abstract}

\section{Introduction}
\label{Sec:Intro}
Many physical systems can be studied using graph models consisting of pairs of nodes connected together via links or edges \cite{Grin14}. 
From information flow in communications and transport infrastructures, to social interactions, biological organisms and semantics, a varied array of systems can all be modelled and studied in terms of complex networks \cite{NewRev} (see Ref.\cite{NewBook}  for an introduction).

\begin{figure}[htb!]
\begin{center}
\includegraphics[width=7cm]{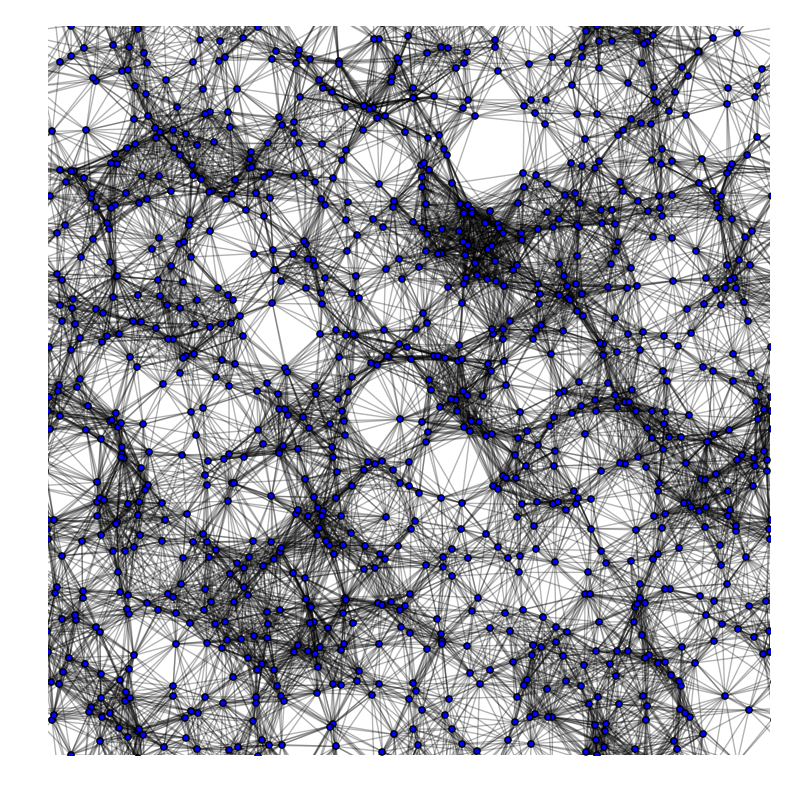}
\caption{A random geometric graph. Here we have illustrated a random geometric graph which consists of $10^3$ nodes uniformly distributed onto the two-dimensional unit torus (blue discs). These nodes are connected by edges (black lines) when they are within a range of $0.1$ of each other. }
\label{fig:RGG}
\end{center}
\end{figure}

One way of studying these systems is to randomly generate or synthesize graph topologies which reproduce the interesting features or structure one is interested in. These models can be studied analytically or ensembles created which can be analysed numerically either directly or fed into larger simulation software packages.
Several random graph models have been created for this purpose such as the Erd\H{o}s-R\'{e}nyi (E-R) random graph model \cite{ErRen59}, the Barab{\'a}si-Albert scale-free network model (B-A)\cite{BarAlb99}, the Watts-Strogatz small-world network model (W-S) \cite{WaSt98} and the random geometric graph (RGG) \cite{Gil61,PenroseBook,WalSurvey} which we focus on here (see figure (\ref{fig:RGG})).

Recently, spectral graph theory has provided the vehicle with which random matrix theory (RMT) can be applied to study statistics of the graph spectrum. Like in traditional spectroscopy, one can then infer structural properties of complex networks. Many types of random graph models have been analysed, however, the ubiquitous and fundamental class of geometric graphs which are the simplest models of spatial networks \cite{Bart11} has yet to be studied using the RMT framework.

\begin{figure}[t!]
\begin{center}
\includegraphics[width=7cm]{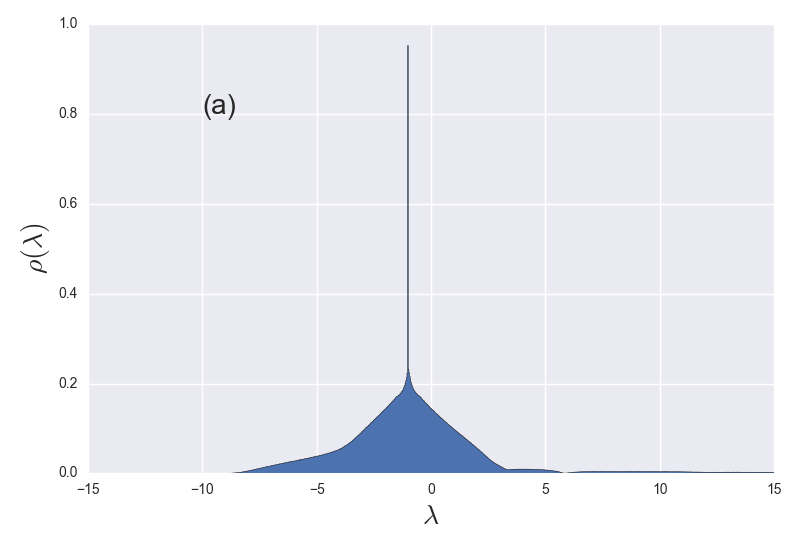}
\includegraphics[width=7cm]{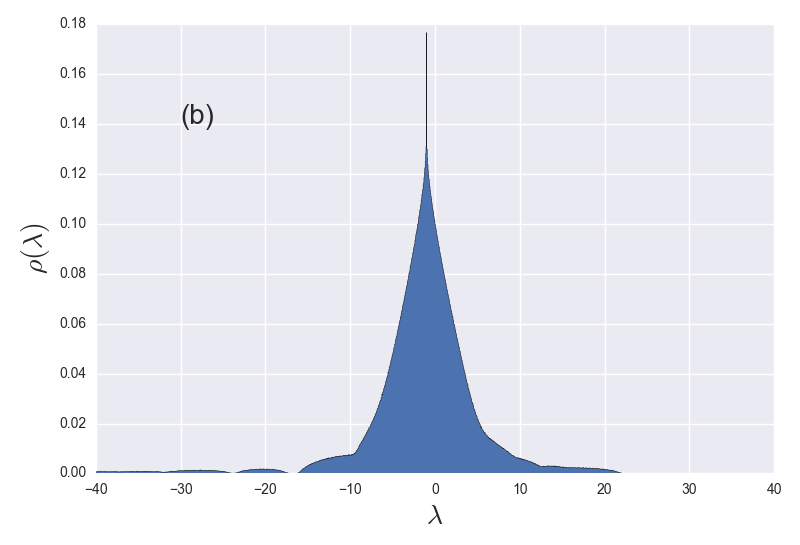}
\caption{Here we illustrate the adjacency matrix spectral density calculated from an ensemble of $10^4$, $10^3$ node RGGs with connection radius $0.1$ (a) and $0.3$ (b). We note the sharp peak in the spectrum at $-1$ caused by the appearance of particular symmetric motifs in RGGs.}
\label{fig:sd}
\end{center}
\end{figure}

A geometric graph is  a spatially embedded network in which all nodes have a well defined location within a given geometric domain.
Thus, geometry structures the network while greatly affecting its connectivity properties.
Indeed, many real-world networks such as transportation networks, the Internet, mobile phone networks, power grids, social networks and neural networks all have a fundamental spatial element to them (see \cite{Bart11} for a survey). 
In this first foray into the spectral properties of geometric graphs using RMT, we specifically focus on the well studied unit-disk RGG model \cite{Gil61,PenroseBook,WalSurvey}.
It is already known that the spectrum of RGGs is very different to the other random graph models mentioned above in that the appearance of particular sub-graphs give rise to multiple repeated eigenvalues \cite{Nyberg2015, BEJ06}. 
This in turn causes sharp peaks to appear in the adjacency matrix spectral density (see figure (\ref{fig:sd})). 
Whilst the appearance of the sharp peaks has been studied, the remaining part of the spectrum remains largely unexplored. 
RMT will allow us to study the spectrum of RGGs and compare with previous results related to other models.

RMT has been applied to a variety of complex networks.
Graph matrices (e.g. adjacency, Laplacian) are first extracted from empirical data or generated from prescribed algorithms.
These are then analysed by looking at the inter-eigenvalue distances (so called level spacings).
In Ref.\cite{LuoEtAl2006} RMT was applied to the study of biological networks where the spectrum of a yeast protein-protein interaction network and a yeast metabolic network were studied. 
Remarkably, the statistics of the level spacings were very similar to those of matrices whose entries are Gaussian distributed random variables; the Gaussian orthogonal ensemble (GOE) statistics of RMT. 
After introducing a modular structure via the removal of particular edges in these biological networks, the level spacing statistics changed from GOE to being Poisson distributed. 
Following this discovery, E-R random graphs were analysed in Ref.\cite{PaVa2006}. 
In E-R random graphs each node is connected to every other with a given probability $p$. 
GOE statistics were observed for highly connected E-R graphs experiencing a transition to Poisson statistics for smaller values of $p$. 
Since these numerical discoveries, a local semi-circle law, which states that the spectral density of GOE matrices is close to Wigner's semicircle distribution on scales containing just more than one eigenvalue, was proven for E-R graphs under the restriction $pN\to\infty$ (with at least logarithmic speed in $N$) \cite{ErdosEtAl2013}.
The latter was also used to prove the presence of GOE statistics in the level spacings of E-R graphs under these conditions \cite{ErdosEtAl2012}.
In fact, the RMT framework has been useful in manifold applications, ranging from differentiating between cancerous and healthy protein networks\cite{Rai2014}, to studying Anderson localisation in complex networks \cite{ZhuEtAl2008,SadeEtAl2005}. 
Further use of RMT in complex networks has focused on the universality properties of these GOE statistics across different random graph models\cite{Jakobson1999,JaBa07,Ja2009,MeEtal15,Riv16}. An overview of the relationship between complex networks ( with specific reference to biological networks) and random matrix theory can be found in Ref. \cite{Ja2015}.
E-R, B-A and W-S have all been studied and similar GOE statistics have been found despite the fact that the spectral densities themselves are very different \cite{FarEtAl2006}.

In this paper we apply for the first time the RMT framework to geometric graphs. 
We first describe the model then provide background to aid in the understanding the RMT framework that we will employ. 
This is subsequently applied numerically to investigate the short-range correlations in the level spacings via the nearest neighbour spacing distribution (NNSD) and the next-nearest neighbour spacing distribution (nNNSD) of the spectra.
These short-range correlation statistics encode information about community structure, connectivity and localisation which has applications to the Anderson metal insulator transition in networks \cite{SadeEtAl2005}. 
We then look at the spectral rigidity in order to investigate the long range correlations of the RGG spectra via the $\Delta_3$ statistic. 
These long-range correlations and the $\Delta_3$ statistic give a measure of the amount of randomness in the connections\cite{Ja2009,JaBa2009}.

\section{Model}
\label{Sec:Model}
In a RGG the nodes are distributed randomly throughout a given domain and the edges are determined by the locations of the nodes, see for example Refs.\cite{PenroseBook}  and \cite{WalSurvey} for introductions. RGGs find particular use in modelling spatial networks such as wireless networks \cite{GupHum99,HaeEtAl2009,Pot2000,EstEtAl99}, epidemic spreading \cite{Wang2009,Nek2007,TorGuc2007}, city growth \cite{Wat2010}, power grids \cite{XiYe2011} and protein-protein interaction networks \cite{HigEtAl2008} for example. There has also been recent interest in studying the properties of RGGs like synchronisation \cite{EstGua2015,DiazEtAl2009}, consensus dynamics \cite{EstShe2016}, connectivity properties \cite{DetOre2016} and spectral properties \cite{Nyberg2015,BEJ06}.

We study RGGs on the unit torus by uniformly distributing $N$ nodes in the unit square and connecting them with an edge when they are within a given range $r$ of each other, using periodic boundary conditions. See figure (\ref{fig:RGG}) for an illustration of a particular realisation with $r=0.1$. We then extract the $N \times N$ adjacency matrix $\mathbf{A}$ of the RGG which has entries $a_{ij}=1$ when there is a connection between nodes $i$ and $j$ and zero otherwise.  $\mathbf{A}$ is a type of Euclidean random matrix which are often studied in random matrix theory (RMT) \cite{Mezard99}. An $N \times N$ Euclidean random matrix has entries $a_{ij}$ which are given by a deterministic function $f(\mathbf{x}_i,\mathbf{x}_j)$ of the locations $\mathbf{x}_i,\mathbf{x}_j$ of  $N$ randomly distributed points. In our RGGs we have
\begin{equation}
f(\mathbf{x}_i,\mathbf{x}_j):=\left\{
        \begin{array}{ll}
          1 & ||\mathbf{x}_i-\mathbf{x}_j || \leq r\\
          0 & ||\mathbf{x}_i-\mathbf{x}_j || > r
        \end{array}
      \right.
\label{Eq:ConnectionFunction}
\end{equation}
The resulting adjacency matrix $\mathbf{A}$ when using Eq.(\ref{Eq:ConnectionFunction}) is real and symmetric hence its spectrum consists of real eigenvalues $\lambda_i, i=1,..,N$ and $\lambda_1 \leq \lambda_2 \leq ... \leq \lambda_N$. We study $\mathbf{A}$ as the spectrum of a network encodes valuable information about the underlying topology \cite{ChungBook}. In Refs. \cite{Nyberg2015} and \cite{BEJ06} it is noted that the ensemble averaged spectral density $\rho(\lambda)$ of RGGs consists of sharp peaks at integer values (in Ref.\cite{Nyberg2015} the related graph Laplacian is studied) caused by the appearance of particular subgraphs whose nodes have the same adjacencies called symmetric motifs (see figure (\ref{fig:sd}) for an illustration of this). This phenomenon is not commonly found in non-spatial network models. In Ref.\cite{Nyberg2015} they refer to these peaks in the spectral density as the discrete part and the remainder as the continuous part. Here we statistically analyse the continuous part of the spectral density using RMT. 

As the parameter $r$ is varied the properties of the RGG change also. On a microscopic scale the mean degree of the nodes is proportional to $r^2$ whilst macroscopically the graph can be disconnected for small $r$ and connected as $r$ increases. As $r$ increases further every node will connect to every other and the RGG becomes the complete graph with trivial spectrum $(N-1)^1, (-1)^{N-1}$. We look at a range of values of $r$ from relatively small ($0.03$) and likely to contain many disconnected components to relatively large ($0.4$) and likely to consist of one connected component  in order to assess how variation of this parameter affects the spectral spacing statistics. See figure \ref{fig:nnsd}.(b) below for how the probability of obtaining a single connected component ($P_{fc}$) depends on $r$.

\section{Random matrix theory}
\label{sec:SpSt}

\begin{figure}[tb!]
\begin{center}
\includegraphics[width=7cm]{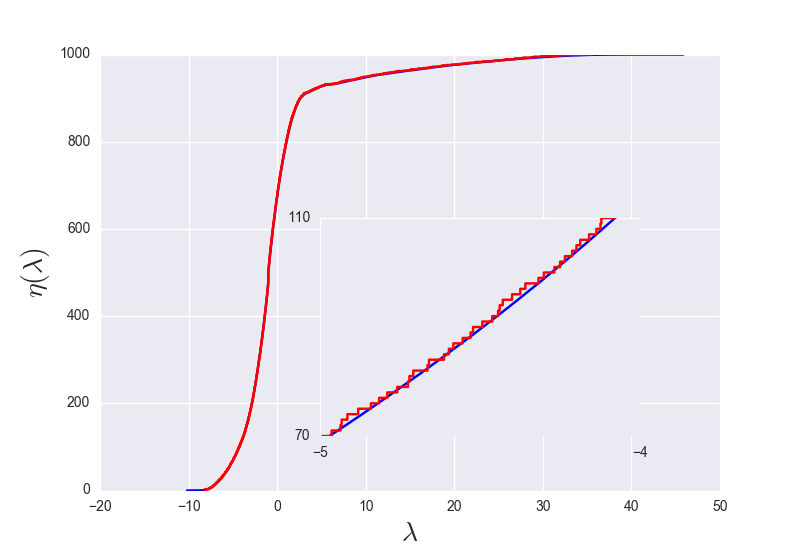}
\caption{Cumulative spectral density. Here the cumulative mean spectral function is illustrated (blue), calculated from an ensemble of $10^4$, $10^3$ node RGGs with connection radius $0.1$ along with the cumulative spectral density of a single RGG (red).}
\label{fig:csd}
\end{center}
\end{figure}

Wigner first developed RMT to study the statistics of eigenvalue spectra of complex quantum systems, see Refs.\cite{Gu98} and \cite{Me04} for reviews and introductions to the subject. It has since been applied to many other types of complex systems \cite{Gu98}. In order to analyse the statistics the spectrum has to be unfolded to create a constant level density \cite{Gu98,Me04}. Examples of the spectral densities which we will be unfolding are illustrated in figure \ref{fig:sd}. To unfold the spectrum we firstly consider the spectral function which for a given {\em energy} $E$ is defined as
\begin{equation}
                           S(E)=\sum_{i=1}^{N}\delta(E-\lambda_i).
\label{Eq:Spectral_fnc}
\end{equation}
The corresponding cumulative spectral function counts how many eigenvalues there are less than or equal to $E$
\begin{equation}
                            \eta(E)=\int_{-\infty}^{E}S(x)dx=\sum_{i=1}^N\Theta(E-\lambda_i).
\label{Eq:CumulSpecFnc}
\end{equation}
The unfolded eigenvalues are then defined in terms of the cumulative mean spectral function
\begin{equation}
                        \overline{\lambda}_i=\langle \eta(E) \rangle |_{E=\lambda_i},
\label{Eq:unfolded}
\end{equation}
where $\langle...\rangle$ signifies a mean value. An analytical form of $\langle \eta(\lambda) \rangle$ is often unobtainable so we use an ensemble average to calculate the mean and then perform the unfolding. See  figure \ref{fig:csd} for an illustration of $\langle \eta(\lambda) \rangle$.

Once a spectrum has been unfolded we can look at the spacing statistics. The nearest neighbour spacings are defined as,
\begin{equation}
                         s_i=\overline{\lambda}_{i+1}-\overline{\lambda}_i.
\label{Eq:nns}
\end{equation}
Due to the unfolding process the expected value $\langle s \rangle$ is unity irrespective of the spectral density $\rho(\lambda)$, but the NNSD $P(s)$ is not unique. For an uncorrelated sequence of points the spacings distribution follows Poisson statistics, i.e.
\begin{equation}
                          P_{po}(s)=e^{-s}.
\label{Eq:Poisson}
\end{equation}
In the case of GOE statistics there are correlations between eigenvalues. A good approximation to the NNSD of GOE matrices is given by the Wigner surmise 
\begin{equation}
                          P_{GOE}(s)\simeq \frac{\pi}{2}se^{-\frac{\pi s^2}{4}}.
\label{Eq:WD}
\end{equation}
Eq.(\ref{Eq:WD}) is exact in the case of $2 \times 2$ matrices and provides a good approximation for larger matrices (see Ref.\cite{Me04} figure $1.5$). The Brody distribution was introduced as a way of interpolating between the two distributions \cite{Brody1973}
\begin{equation}
                          P_{\beta}(s)=(\beta +1)\alpha s^{\beta}e^{-\alpha s^{\beta +1}},
\label{Eq:Brodydist}
\end{equation}
where
\begin{equation}
                          \alpha=\Gamma\left( \frac{\beta +2}{\beta +1}\right)^{\beta +1},
\label{Eq:alpha}
\end{equation}
$\Gamma()$ is the Gamma function.  $\beta=0 $ corresponds to the Poisson statistics Eq.(\ref{Eq:Poisson}) whilst $\beta=1$ to the Wigner surmise Eq.(\ref{Eq:WD}). We stress that there is no physical significance to the parameter $\beta$ in the Brody distribution but it has been noted that it captures the transition from Poisson to GOE statistics rather well \cite{Che90}. Furthermore the Brody distribution is frequently used in the study of complex networks to measure the transition between and mixture of GOE and Poisson statistics \cite{PaVa2006,ZhuEtAl2008,JaBa07,Ja2009,MeEtal15,JaBa072}. Hence we use it here for comparison. 

\section{Nearest neighbour spacings}
\label{sec:nnsd}
\begin{figure}[hbt]
\begin{center}
\includegraphics[width=8.5cm]{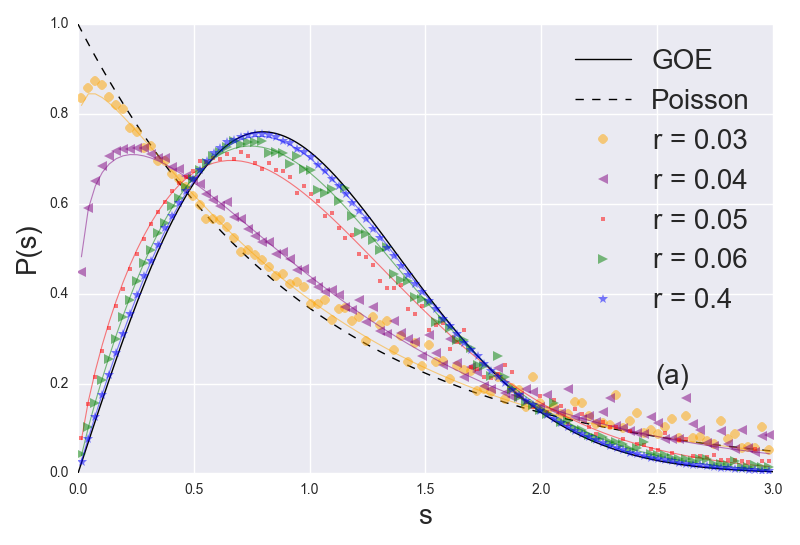}\\ \includegraphics[width=8.5cm]{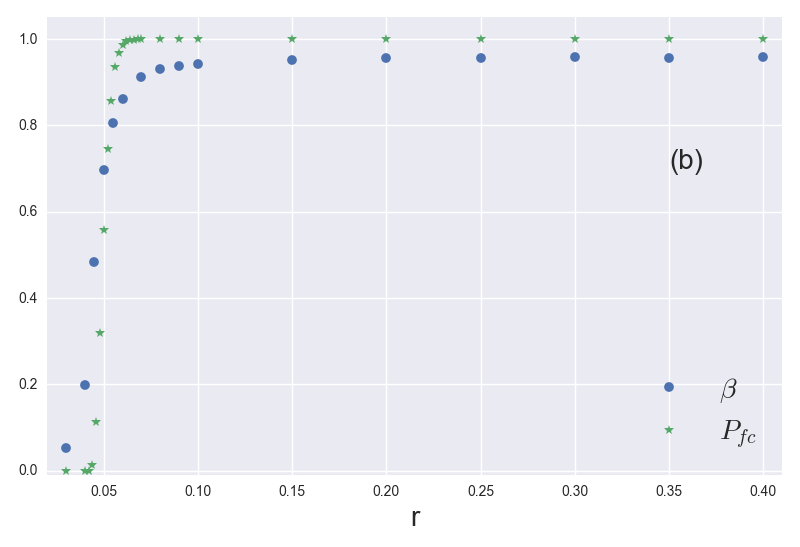}
\caption{Nearest neighbour spacings of unfolded eigenvalues. Here the NNSD is numerically calculated from an ensemble of $10^4$, $10^3$ node RGGs and illustrated for a range of connection values in (a) along with the Brody distribution fit (lines) and the NNSD for Poisson and GOE statistics. In (b) we show the best fit parameter $\beta$ to the NNSD for a range of $r$ values showing the transition from Poisson ($\beta = 0$) to GOE ($\beta = 1$) (blue dots) along with the probability of full connectivity $P_{fc}$ calculated from ensembles of $10^4$ RGGs (green stars).}
\label{fig:nnsd}
\end{center}
\end{figure} 

We calculated the NNSD $P(s)$ from an ensemble of RGGs at various values of the connection radius $r$. To obtain $P(s)$ we firstly calculate the spectrum of an individual RGG. This is then unfolded to remove the system specific effects and the $s_i$ are extracted. This process is performed for an ensemble of RGGs to obtain $P(s)$, see Ref.\cite{Shriner1992} for an error analysis of these statistics. We then fit the Brody distribution of Eq.(\ref{Eq:Brodydist}) to $P(s)$ and interpret the fit parameter $\beta$ as a measure of similarity to either GOE or Poisson statistics. 

We firstly note that there appears a sharp peak at zero in the NNSD of RGGs. This is not due to a degeneracy caused by disconnected components, as it appears for connected RGGs. Rather this is caused by the multiplicity of $-1$ in the spectrum as discussed earlier (figure \ref{fig:sd}(a)).  We remove this peak and calculate the NNSD. This is illustrated for a range of $r$ values in figure \ref{fig:nnsd}(a) along with the Brody distribution fit. Table \ref{Table:NNSDtable} contains the standard error of the best fit estimate along with the $\chi^2$ statistic. For small values of $r$ the mean degree of the vertices is also relatively low. At $r=0.03$ the mean degree is less than three. Hence it is highly likely that the RGGs consist of many isolated components (communities) and the spectrum will consist of the union of independent spectra. Correspondingly we see very few correlations in the NNSD illustrated by low $\beta$ at low $r$ values. As $r$ increases the mean degree increases quadratically. The isolated components merge until the graph consists of a single connected component. The probability of obtaining a fully connected RGG at a given $r$ value ($P_{fc}$) was calculated numerically and is also illustrated in figure \ref{fig:nnsd}(b). We see that as $P_{fc}$ transitions from zero to one we observe a transition from Poisson to GOE statistics in the NNSD.

In Ref. \cite{LuoEtAl2006} GOE statistics in the NNSD of a complex network is interpreted as indicative of a lack of modular or community structure, Poisson statistics being found in highly modular networks. Furthermore the NNSD is also studied in terms of the Anderson metal-insulator transition of localised to extended eigenstates in complex networks. GOE statistics are characteristic of extended eigenstates whilst Poisson statistics indicate localisation \cite{SadeEtAl2005}. In RGGs for small $r$ the eigenstates will be localised on the disconnected components.

\begin{table}[h]
\begin{center}
\begin{tabular}{|c||c|c||c|c|}
\hline
$r$ & $\beta$ & $\chi^2$& $KS$ value& $p$ value\\
\hline
0.03    & $0.052 \pm 0.005$ & $0.064$  & $0.192$ & $0.000$  \\
0.04    & $0.198 \pm 0.006$ & $0.050$  & $0.151$ & $0.000$  \\
0.05    & $0.696 \pm 0.008$ & $0.029$  & $0.060$ & $0.000$  \\
0.06    & $0.862 \pm 0.006$ & $0.013$  & $0.031$ & $0.000$  \\
0.07    & $0.912 \pm 0.005$ & $0.010$  & $0.023$ & $0.000$  \\
0.08    & $0.931 \pm 0.004$ & $0.007$  & $0.014$ & $0.000$  \\
0.09    & $0.937 \pm 0.004$ & $0.006$  & $0.010$ & $0.000$ \\
0.1      & $0.942 \pm 0.004$ & $0.005$  & $0.008$ & $0.000$ \\
0.2      & $0.955 \pm 0.002$ & $0.002$  & $0.004$ & $0.155$ \\
0.3      & $0.957 \pm 0.002$ & $0.002$  & $0.001$ & $0.989$ \\
0.4      & $0.958 \pm 0.002$ & $0.001$  & $0.002$ & $0.916$ \\
\hline
\end{tabular}
\caption{ In this table is the best parameter fit for $\beta$ of Eq.(\ref{Eq:Brodydist}) to the numerically obtained nearest neighbour spacing distribution as a function of connection radius $r$ along with the standard error and corresponding  $\chi^2$ statistic. Also reported is the Kolmogorov-Smirnov statistic of the numerically obtained next nearest neighbour spacing distribution tested against Eq.(\ref{Eq:GSE}) along with the corresponding $p$ value. }
\label{Table:NNSDtable}
\end{center}
\end{table}

\begin{figure}[ht!]
\begin{center}
\includegraphics[width=8.5cm]{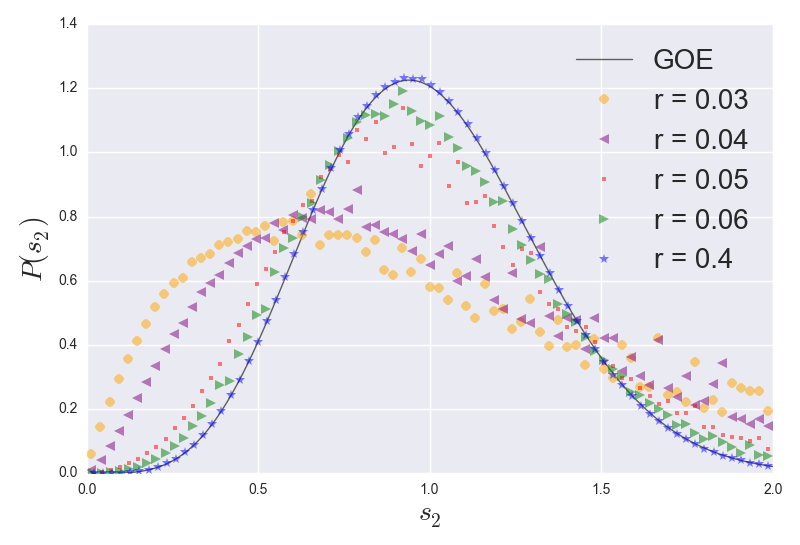}
\caption{Next nearest neighbour spacings of unfolded eigenvalues. Here the nNNSD $P(s_2)$ is calculated from an ensemble of $10^4$, $10^3$ node RGGs for a range of connection values. Also illustrated is the nNNSD for GOE statistics.}
\label{fig:nnnsd}
\end{center}
\end{figure}

An additional statistic used to study complex networks \cite{JaBa07} is the {\em next} nearest neighbour spacings of the unfolded eigenvalues $s_2$ where 
\begin{equation}
                        s_2^i=   (\overline{\lambda}_{i+2}-  \overline{\lambda}_i)/2,
\label{Eq:ps2}
\end{equation}
and their distribution $P(s_2)$. The factor of two in Eq.(\ref{Eq:ps2}) again ensures a mean spacing of unity. The nNNSD of the GOE is given by the NNSD of the Gaussian symplectic ensemble of random matrices (GSE) which is well approximated by (see Ref.\cite{Me04})
\begin{equation}
                         P_{GSE}(s_1)\simeq\frac{2^{18}}{3^6\pi^3}s_1^4e^{-\frac{64}{9\pi}s_1^2}.
\label{Eq:GSE}
\end{equation}
We similarly calculated $P(s_2)$  for an ensemble of RGGs which can be seen in figure \ref{fig:nnnsd}. We again observed a peak at zero caused by the discrete peak in the spectral density. After removal of this peak we see that the nNNSD of RGGs fits very closely to that of the GOE statistics for large $r$ (well connected) given by Eq.(\ref{Eq:GSE}) but we observe a transition away from this as $r$ is decreased and the RGGs become disconnected. Table \ref{Table:NNSDtable} captures this transition via the Kolmogorov-Smirnov statistic where we observe a sharp drop in the $p$ value between $0.3$ and $0.2$. GOE statistics have been found in the nNNSD of  $N=2000$ mean degree $20$ (connected) non-spatial (E-R, scale-free and small-world) networks \cite{JaBa07}.
\section{Spectral rigidity}
\label{sec:SR}

\begin{figure}[ht!]
\begin{center}
\includegraphics[width=8.5cm]{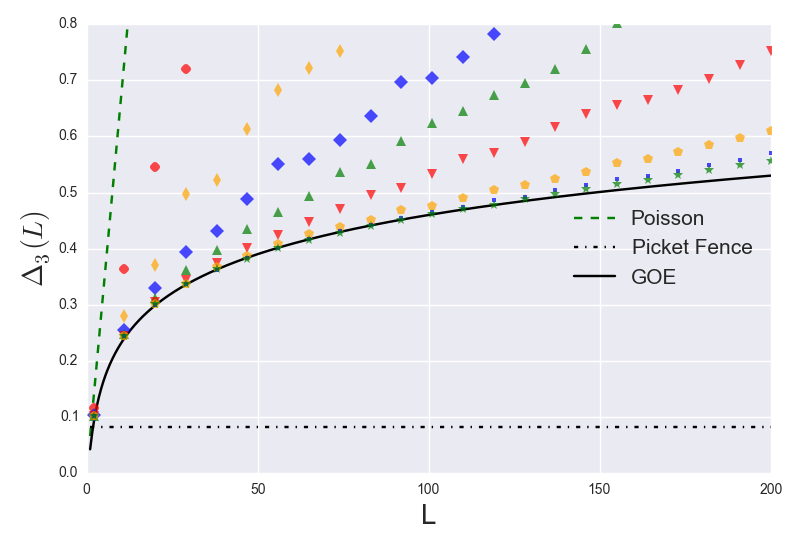}
\caption{Spectral rigidity of RGGs. Here is illustrated the spectral rigidity, calculated from an ensemble of $10^3, 10^3$ node RGGs with $r=0.05, 0.06, 0.07, 0.08, 0.1, 0.15, 0.2,  0.4$ (red circles, orange thin diamonds, blue diamonds, green triangles (up facing), red triangles (down facing), orange pentagons, blue dots, green stars respectively). Also illustrated is the result predicted by GOE statistics (black line), Poisson statistics (green dashed line) and the even spacing of the {\em picket fence} spectrum (dot-dash black line).}
\label{fig:SR}
\end{center}
\end{figure}

So far we have only looked at short range correlations in the spectra via the NNSD and nNNSD. We will now look at the $\Delta_3$ statistic, introduced in Ref.\cite{DyMe63}, which measures long range correlations. $\Delta_3(L,x)$ measures the least-square deviation of the unfolded spectral staircase function $\overline{\eta}$ to the line of best fit over the interval $[x,x+L]$.
\begin{equation}
                        \Delta_3(L,x)=\frac{1}{L}\min_{A,B} \int_x^{x+L}\left( \overline{\eta}(\overline{\lambda})-A\overline{\lambda}-B\right)^2 d\overline{\lambda}.
\label{Eq:DeltaLx}
\end{equation}
 Where $ \overline{\eta}$ counts how many unfolded eigenvalues there are less than or equal to a given value
\begin{equation}
                            \overline{\eta}(E)=\sum_{i=1}^N\Theta(E-\overline{\lambda}_i).
\label{Eq:CumulSpecFncUNF}
\end{equation}
The average over non-intersecting intervals of length $L$ $\langle...\rangle_x$ is then the spectral rigidity $\Delta_3(L)$. 
\begin{equation}
                          \langle  \Delta_3(L,x) \rangle_x=\Delta_3(L).
\label{Eq:DeltaxAv}
\end{equation}
For full correlation where all the spacings are equal, such as that of the harmonic oscillator, the so-called {\em picket fence} spectrum there is no dependence on $L$
\begin{equation}
                        \Delta_3(L)=\frac{1}{12}.
\label{Eq:DeltaPoiss}
\end{equation}
Meanwhile, a fully uncorrelated random sequence gives Poisson statistics in the spacings. In this case there is linear dependence on $L$ given by 
\begin{equation}
                        \Delta_3(L)=\frac{L}{15}.
\label{Eq:DeltaPoiss}
\end{equation}
GOE statistics sit in between these two cases with a logarithmic dependence on $L$. For large $L$
\begin{equation}
                        \Delta_3(L)\simeq \frac{1}{\pi^2}\left( \ln(2\pi L)+\gamma-\frac{5}{4}-\frac{\pi^2}{8} \right),
\label{Eq:DeltaLo}
\end{equation}
to order $1/L$ \cite{DyMe63}, where $\gamma$ is Euler's constant. A useful technique for evaluating  $\Delta_3(L,x)$ has been developed in \cite{Boh75} and outlined in \cite{Boh84} for an experimentally obtained sequence. This involves first shifting the interval $[x,x+L]$ so that its centre is at the origin, i.e. for all the unfolded eigenvalues $\overline{\lambda}_i,\overline{\lambda}_{i+1},...,\overline{\lambda}_{i+n-1}$ we shift them (and relabel for convenience)
\begin{equation}
                       \hat{\lambda}_j=\overline{\lambda}_{i-1+j}-\left(x+\frac{L}{2}\right),
\label{Eq:EigenTrans}
\end{equation}
we then have the following
\begin{eqnarray}\nonumber
                        \Delta_3(L,x) = \frac{n^2}{16}-\frac{1}{L^2}\left( \sum_{j=1}^n  \hat{\lambda}_j\right)^2
+\frac{3n}{2L^2}\left( \sum_{j=1}^n  \hat{\lambda}_j^2\right) \\
-\frac{3}{L^4}\left( \sum_{j=1}^n  \hat{\lambda}_j^2\right)^2
+\frac{1}{L}\left( \sum_{j=1}^n  (n-2j+1)\hat{\lambda}_j\right).
\label{Eq:DeltaFformula}
\end{eqnarray}
Using Eq.(\ref{Eq:DeltaFformula}) we evaluate $\Delta_3(L)$, being careful not to sample the discrete peaks in the spectral density (this creates large jumps in the staircase function). See figure \ref{fig:SR} for an illustration of $\Delta_3(L)$ for a range of $r$ values. We see that the RGGs follow the GOE statistics up to some value $L_0$ and then deviate towards Poisson statistics, with the value of $L_0$ depending on $r$. The larger $r$ gives larger $L_0$. In Ref.\cite{JaBa07} they find very good agreement between the $\Delta_3$ statistic of the E-R random networks they study and the GOE statistic for large values of $L$, which is to be expected given the results in Refs.\cite{ErdosEtAl2013} and \cite{ErdosEtAl2012} on the similarity between GOE and well connected E-R graphs. Whilst for the scale-free and small-world networks they find good agreement up to certain values of $L$ but then they see deviations towards Poisson statistics as we have  observed here in RGGs. Indeed in Ref.\cite{Ja2009} they show how the value of $L_0$ is related to the amount of community structure within the network by analysing networks constructed from randomly connected E-R networks. Furthermore in  Ref.\cite{JaBa2009} the value $L_0/N$ is interpreted as a measure of the amount of randomness in the connections of the network. This amount of randomness is defined in terms of the randomness introduced via a rewiring probability in regular degree networks. The higher the rewiring probability the larger $L_0$.
\section{Summary}
\label{sec:summary}
Here we have numerically analysed the spectrum of the adjacency matrices of spatial networks by looking at the random geometric graph model using a random matrix theory framework. We analysed two statistics which look at short-range correlations in the level spacings of the spectrum; the nearest neighbour distribution and the  next nearest neighbour distribution. We also analysed the spectral rigidity via the $\Delta_3$ statistic which looks at long-range correlations. These statistics give insight into localisation, community structure and randomness in complex networks. 

Firstly we found that the relatively common appearance of certain symmetric motifs in random geometric graphs appear as a peak at $0$ in the nearest neighbour distributions. We also found that despite the deterministic connection function used (Eq.(\ref{Eq:ConnectionFunction})) random geometric graphs are statistically very similar to certain types of random graph which have been studied like the Erd\H{o}s-R\'{e}nyi random graphs, Barab{\'a}si-Albert scale-free networks and the Watts-Strogatz small-world networks \cite{JaBa07} in that the statistics display a parameter dependent transition between the Gaussian orthogonal ensemble of random matrices for high $r$ values and closer to Poisson statistics for low $r$ values. In terms of network structure these results are indicative of the connectivity transition from many isolated components at low $r$ values to a single connected component at high values of $r$. This transition has also been interpreted in terms of the level of randomness in the connections of random graphs \cite{JaBa2009}. Furthermore in terms of Anderson localisation it is seen in the transition from localised to delocalised eigenstates \cite{SadeEtAl2005}.

The connection function we have studied given by Eq.(\ref{Eq:ConnectionFunction}) is fundamental to the study of random geometric graphs \cite{PenroseBook} but there are other, more general, random connection functions that one can study \cite{DetOre2016}. Future work will investigate these connection functions and look at how the additional randomness is reflected in particular in the $\Delta_3$ statistic.  For this it will also be important to capture the transition between and mixing of random Poisson and correlated Gaussian orthogonal ensemble statistics. We saw how this transition was captured by the often used Brody distribution Eq.(\ref{Eq:Brodydist}) so this could possibly provide a good starting point. Generalising the results in Refs.\cite{ErdosEtAl2013} and \cite{ErdosEtAl2012} could also potentially give analytical answers to these questions. Furthermore it will be interesting to study the spectral properties of other types of networks such as self-similar  \cite{Gallos12} or even multiplex networks \cite{Yanqing14,Danzinger16} using RMT.

\section*{Acknowledgements}
We would like to thank Justin P. Coon, Sarika Jalan and Jonathan P. Keating for helpful discussions and comments. This work was supported by the EPSRC grant number EP/N002458/1 for the project Spatially Embedded Networks.


\end{document}